\documentclass[10pt]{article}
\usepackage{graphicx}
\usepackage{latexsym,graphicx}
\usepackage{float}
\usepackage{amsmath}
\usepackage{amsfonts}
\usepackage{amssymb}
\usepackage{epstopdf}
\DeclareGraphicsExtensions{.eps}
\usepackage[left=2.5cm,top=2.5cm,right=2.5cm,bottom=2.5cm]{geometry}

\begin{document}
	\begin{center}
		\large{\bf{Transit cosmological models with observational constraints in $f(Q, T)$ gravity}} \\
		\vspace{5mm}
		\normalsize{ Anirudh Pradhan$^1$, Archana Dixit$^2$}\\
		\vspace{5mm}
		\normalsize{$^{1,2}$Department of Mathematics, Institute of Applied Sciences \& Humanities, GLA University,\\ 
                 Mathura -281 406, Uttar Pradesh, India} \\
                 \vspace{5mm}
                $^1${Email:pradhan.anirudh@gmail.com}\\
                \vspace{2mm}
                $^2${Email:archana.dixit@gla.ac.in}\\
	\end{center}
	\vspace{10mm}
	%\date{}
	%\maketitle
	%%%%%%%%%%%%%%%%%%%%%%%%%%%%%%%%%%%%%%%%%%%%%%%%%%%%%%%%%%%%%%%%%%%%%%%%%%%%%%%%%%%%%%%%%%%%
	\begin{abstract}
This cosmological model is a study of modified $f(Q,T)$ theory of gravity which was recently proposed by Xu {\it et al.} (Eur. Phys. J. C {\bf 79}, 708 (2019)). 
In this theory of gravity, the action contains an arbitrary function $f(Q,T)$ where $Q$ is non-metricity and $T$ is the trace of energy-momentum tensor 
for matter fluid. In our research, we have taken the function $f(Q,T)$ quadratic in $Q$ and linear in $T$ as $f(Q,T)=\alpha Q+\beta Q^{2}+\gamma T$ 
where $\alpha$, $\beta$ and $\gamma$ are model parameters, motivated by $f(R,T)$ gravity.  We have obtained the various cosmological parameters in 
Friedmann-Lemaitre-Robertson-walker (FLRW) Universe viz. Hubble parameter $H$, deceleration parameter $q$ etc. in terms of scale-factor as well as in 
terms of redshift $z$ by constraining energy-conservation law.  For observational constraints on the model, we have obtained the best-fit values of model 
parameters using the available data sets like Hubble data sets $H(z)$, Joint Light Curve Analysis (JLA) data sets and union $2.1$ compilation of SNe Ia data 
sets by applying $R^{2}$-test formula. We have calculated the present values of various observational parameters viz.  $H_{0}$, $q_{0}$, $t_{0}$ and statefinder 
parameters $(s,r)$, these values are very close to the standard cosmological models. Also, we have observed that the deceleration parameter $q(z)$ shows 
signature-flipping (transition) point within the range $0.423\leq z_{t}\leq0.668$ through which it changes its phase from decelerated to accelerated 
expanding universe with equation of state (EoS) $-1.071\leq\omega-0.96$ for $0\leq z\leq3$.			
	\end{abstract}
	\smallskip
	\vspace{5mm}
	%\date{}
	%\maketitle
	\vspace{1cm}
	\smallskip
	%%%%%%%%%%%%%%%%%%%%%%%%%%%%%%%%%%%%%%%%%%%%%%%%%%%%%%%%%%%%%%%%%%%%%
	%%%%%%%%%%%%%%%%%%%%%%%%%%%%%%%%%%%%%%%%%%%%%%%SECTION 1
	%%%%%%%%%%%%%%%%%%%%%%%%%%%%%%%%%%%%%%%%%%%%%%%%%%%%%%%%%%%%%%%%%%%%%%%%%%%%%%%%%%%%%%%%%%%%%%%%%%%%%%%
	\section{Introduction}
Even if, Einstein's general relativity (GR) proposed in $1915$ is still considered to be one of the most successful theories, but in the wake of 
new observational advances in  cosmology, there appears some limitations on standard GR in explaining those phenomena. The observations in 
\cite{ref1}-\cite{ref7} have strong evidence that our present universe is undergoing an accelerated expansion phase. These same observations predict 
the existence of two unknown components in the Universe, called dark matter (DM) and dark energy (DE) along with the usual baryonic matter. The heavy 
negative pressure due to DE is supposed to be the possible cause of the accelerated expansion. \\

To adapt new findings, there are mainly two ways, one to provide alternative theories of gravity and the other to modify the Einstein's GR. Theories 
like Brans and Dicke \cite {ref8} , Nordtvedt \cite {ref9}, Wagoner \cite {ref10}, Dunn \cite {ref11}, Saez and Ballester \cite {ref12}, Barber \cite {ref13} 
are some of the prominent alternating theories. To understand the dark energy problem in a better way, time-to-time various cosmologists have proposed some 
well-known modified theories of gravitation viz. $f(G)$ gravity \cite {ref14}, $f(R)$ gravity \cite {ref15}, $f(R,T)$ gravity \cite {ref16}, $f(\mathcal{T})$ 
gravity \cite {ref17}, $f(G,T)$ \cite {ref18},  $f(R, T, R_{\mu\nu} T^{\mu\nu})$ gravity \cite {ref19}, $f(Q)$ theory \cite {ref20} and recently proposed $f(Q,T)$ 
gravity \cite {ref21}.\\

\indent In standard GR the action for formulation of Einstein field equations is given by $S=\int (\frac{R}{2\kappa}+\mathcal{L}_m)\sqrt{-g}d^{4}x$, 
here $g=det(g_{\mu \nu})$ is defined the determinant of metric tensor $g_{\mu\nu}$, $R$ is the Ricci scalar-curvature, $\mathcal{L}_m$ is the matter 
Lagrangian and $\kappa =8\pi G c^{-4}$ is Einstein's constant. In the attempt to modify Einstein's GR, one of the simplest way is to replace $R$ by some 
more general entity. A modified theory of gravity is formulated by replacing the Ricci scalar-curvature $R$ by an arbitrary function $f(R)$ in the 
Einstein–Hilbert action, called as $f(R)$ gravity has proposed by Nojiri and Odintsov \cite{ref15}. The different cosmological features of $f(R)$ gravity 
have been studied by several researchers in different perspectives \cite{ref22}-\cite{ref32}. An extension of $f(R)$ gravity is proposed by Harko 
et al. \cite{ref17} by including the trace of energy-momentum tensor $T_{ij}$ in $f(R)$ function called as $f(R,T)$ gravity. The theory of $f(R,T)$ gravity 
has been rigorously studied by several authors. The references \cite{ref33}-\cite{ref37} are a small list of the study on $f(R,T)$ gravity.\\

Einstein' GR is basically a geometric theory based on Reimannian geometry. In one of the classical approach to modify GR is to a apply some more 
general geometric structures that could describe the gravitational field. The first attempt in this approach is due to Weyl \cite{ref38}, but the main aim 
of the Weyl geometry was to unify electromagnetism and gravitation. On the similar line, Gauss-Bonnet gravity, or $f(G)$ gravity is an interesting modified 
theory, in which gravitation Lagrangian is obtained by adding a generic function $f(G)$ in the Hilbert-Einstein action, where 
$G=R_{\alpha \beta \mu \nu}R^{\alpha \beta \mu \nu}-4R_{\alpha \beta}R^{\alpha \beta}+R^2$ is the Gauss-Bonnet invariant 
($R_{\alpha \beta \mu \nu}$, $R_{\alpha \beta}$ and $R$ are Riemann tensor, Ricci tensor and Ricci scalar respectively) \cite{ref14}. $f(G)$ gravity is further 
studied by many researchers. See \cite{ref39}-\cite{ref43}. Sharif and Ikram \cite{ref18} presented an extension of $f(G)$ gravity theory named as $f(G,T)$ 
gravity by including the trace of the energy-momentum tensor $T$ along with $G$ in the modified  Hilbert-Einstein action. For more study on $f(G,T)$ gravity 
see references \cite{ref44,ref45}  \\

In another approach, the so called teleparallel equivalent to general relativity (TEGR), the basic idea is to replace the metric tensor $g_{\mu \nu}$ 
of the space-time, by a set of tetrad vectors $e^{i}_{\mu}$. The curvature is replaced by the torsion generated by the tetrad, which can be used to entirely 
describe gravitational effects \cite{ref17,ref46,ref47,ref48}. The teleparallel or $f(\mathcal{T})$ gravity are extensively used to explain the current cosmic 
accelerated expansion without using the concept of dark energy \cite{ref49}-\cite{ref60}. \\

In symmetric teleparallel gravity (STG), the basic geometric variable describing the properties of the gravitational interaction is represented by 
the non-metricity  $Q$ of the metric. The non-metricity tensor is the covariant derivative of the metric tensor which geometrically describes the variation 
of the length of a vector in the parallel transport. This approach is first introduced by Nester and Yo in (1999) \cite{ref20}. The STG was further 
developed as $f(Q)$ gravity theory or non-metric gravity and various geometrical and physical aspects of it have been studied in past few years. In the study 
of the cosmology of the $f(Q)$ theory, it has been found that the accelerated expansion is the intrinsic property of the Universe without need of either exotic 
dark energy or extra field \cite{ref61}-\cite{ref75}.\\

 Recently, the $f(Q)$ theory is further extended by Xu {\it et al.} \cite{ref21} in the form of $f(Q,T)$ theory by coupling the non-metricity $Q$ with the 
trace of the matter energy-momentum tensor $T$. The Lagrangian of the gravitational field is assumed to be general function of both $Q$ and $T$. The action 
in $f(Q,T)$ theory is taken in the form $S=\int{\left(\frac{1}{16\pi}f(Q, T)+\mathcal{L}_m\right) \sqrt{-g} d^{4}x}$. The field equations of the theory were 
obtained by varying the gravitational action with respect to both metric and connection. They investigated the cosmological implications of the theory and 
obtained the cosmological evolution equations for a flat, homogeneous, isotropic geometry, by assuming some simple functional form of $f(Q,T)$. 
Arora et al. have recently studied $f(Q,T)$ gravity models with observational constraints \cite{ref76}.\\

In the present study, we have also interpreted the cosmological model with Friedmann-Lemaitre-Robertson-walker (FLRW) Universe in $f(Q,T)$ theory 
with $f(Q,T)=\alpha Q + \beta Q^{2}+\gamma T$ and this supported by $f(R,T)$-gravity form $f(R,T)=R+\alpha R^{2}+\beta T$ in which the 
presence of square term of $R$ reveals the existence of dark energy and dark matter. Using the field equations in a different approach, applying the energy 
conservation condition $\dot{\rho}+3H(\rho + p)=0$, we have obtained the various cosmological parameters viz. Hubble parameter $H(z)$, deceleration parameter 
$q(z)$ etc. in terms of scale-factor $a(t)$ as well as in terms of redshift $z$. Applying $R^2$-test formula and using the available observational data sets 
Hubble data sets \cite{ref77}-\cite{ref83}, Joint Light Curve Analysis (JLA) data sets \cite{ref84} and union $2.1$ compilation of supernovae type Ia data sets 
\cite{ref85} the best fit values of the model parameters are determined. We have obtained the energy density $\rho$ and isotropic pressure $p$ explicitly and 
then estimated the equation of state parameter $\omega=\frac{p}{\rho}$.\\

\indent The present paper is investigated in the following five sections: first section contains a brief introduction, in Sect. $2$ some preliminary 
definitions and an overview of the $f(Q,T)$ theory are given. Cosmological solutions of the FLRW Universe in $f(Q,T)$ theory is given in Sect. $3$. 
Result analysis and discussions are given in Sect. $4$. Finally, conclusions are summarized in Sect. $5$.
%%%%%%%%%%%%%%%%%%%%%%%%%%%%%%%%%%%%%%%%%%%%%%%%%%%%%%%%%%%%%%%%%%%%%%%%%%%%%%
%%%%%%%%%%%%%%%%%%%%%%%%%%%%%%%%%%%%%%%%%%%%%%%%%%%%%%%%%%%%SECTION 2
%%%%%%%%%%%%%%%%%%%%%%%%%%%%%%%%%%%%%%%%%%%%%%%%%%%%%%%%%%%%%%%%%%%%%%%%%%%%%%%%%
\section{Some Preliminary definitions and Cosmological Field Equations}
The Einstein-Hilbert action for $f(Q,T)$ gravity is defined in \cite{ref21} as
\begin{equation}\label{1}
I=\int{\left(\frac{1}{16\pi}f(Q, T)+L_{m}\right)d^{4}x \sqrt{-g}}
\end{equation}
where $f(Q,T)$ is an arbitrary function of the non-metricity $Q$ and the trace of the energy-momentum tensor $T$, and $L_{m}$ is the Lagrangian for 
matter source and $g=\text{det}(g_{ij})$ read as the determinant of the metric tensor $g_{ij}$ and
\begin{equation}\label{2}
Q\equiv-g^{ij}(L^{\alpha}_{\beta i}L^{\beta}_{j\alpha}-L^{\alpha}_{\beta\alpha}L^{\beta}_{ij})
\end{equation}
where $L^{\alpha}_{\beta\gamma}$ is known as deformation tensor defined by
\begin{equation}\label{3}
L^{\alpha}_{\beta\gamma}=-\frac{1}{2}g^{\alpha\lambda}(\nabla_{\gamma}g_{\beta\lambda}+\nabla_{\beta}g_{\lambda\gamma}-\nabla_{\lambda}g_{\beta\gamma})
\end{equation}
The non-metricity $Q$ and the energy-momentum tensor are defined respectively as
\begin{equation}\label{4}
Q_{\alpha}={Q_{\alpha}^{i}}_{i},~~~~T_{ij}=-\frac{2}{\sqrt{-g}}\frac{\delta (\sqrt{-g}L_{m})}{\delta g^{ij}}
\end{equation}
and
\begin{equation}\label{5}
\theta_{ij}=g^{\alpha\beta}\frac{\delta T_{\alpha\beta}}{\delta g^{ij}}
\end{equation}
By varying the action $(1)$ with respect to the metric components, we obtain the following field equations:
\begin{equation}\label{6}
-\frac{2}{\sqrt{-g}}\nabla_{\alpha}(f_{Q}\sqrt{-g}P^{\alpha}_{ij})-\frac{1}{2}f g_{ij}+f_{T}(T_{ij}+\theta_{ij})-
f_{Q}(P_{i\alpha\beta}Q_{j}^{\alpha\beta}-2{Q^{\alpha\beta}}_{i}P_{\alpha\beta j})=8\pi T_{ij}
\end{equation}
where $P^{\alpha}_{ij}$ is defined as the supper-potential of the model mentioned as in \cite{ref21}.\\

Now, let us suppose the universe is isotropic and spatially flat and hence, it can be described by an FLRW metric given by
\begin{equation}\label{7}
ds^{2}=-N^{2}(t)dt^{2}+a(t)^{2}(dx^{2}+dy^{2}+dz^{2})
\end{equation}
where $a(t)$ is called scale factor which describes evolution of the universe. The increasing value of $a(t)$ reveals the expansion and decreasing 
value indicates the collapse of the universe and its rate is measured by Hubble function defined as $H=\frac{\dot{a}}{a}$. The non-metricity $Q$ for 
FLRW metric is derived as $Q=6H^{2}$ for lapse function $N(t)=1$ \cite{ref21}. Let us suppose that our universe is filled with perfect fluid and hence, 
the energy momentum tensor for perfect fluid is defined by $T_{ij}=(\rho+p)u_{i}u_{j}+pg_{ij}$ and therefore, the $\theta_{ij}$ is defined and obtained as
\begin{equation}\label{8}
\theta^{i}_{j}=\delta^{i}_{j}p-2T^{i}_{j}=\text{diag}(2\rho+p, -p, -p, -p)
\end{equation}
Let us introduce the following notations $F\equiv f_{Q}$ and $8\pi \tilde{G}\equiv f_{T}$ for simplicity of field equations. For the FLRW metric $(7)$, 
the derived field equation $(6)$ gives the following equations:
\begin{equation}\label{9}
8\pi\rho=\frac{f}{2}- 6FH^{2}-\frac{2\tilde{G}}{1+\tilde{G}}(\dot{F}H+F\dot{H})
\end{equation}
\begin{equation}\label{10}
8\pi p=-\frac{f}{2}+6FH^{2}+2(\dot{F}H+F\dot{H})
\end{equation}

To obtain the evolution function for Hubble parameter, we have combined the Eqs. $(9)$ and $(10)$ as
\begin{equation}\label{11}
\dot{H}+\frac{\dot{F}}{F}H=\frac{4\pi}{F}(1+\tilde{G})(\rho+p)
\end{equation}
In generalized $f(Q,T)$ gravity the energy-conservation law is derived in \cite{ref21} as
\begin{equation}\label{12}
\dot{\rho}+3H(\rho+p)=\frac{\tilde{G}}{16\pi(1+\tilde{G})(1+2\tilde{G})}\left[\dot{S}-\frac{(3\tilde{G}+2)\dot{\tilde{G}}}{(1+\tilde{G})\tilde{G}}S+6HS\right] 
\end{equation}
where $S=2(\dot{F}H+F\dot{H})$.
%%%%%%%%%%%%%%%%%%%%%%%%%%%%%%%%%%%%%%%%%%%%%%%%%%%%%%%%%%%%%%%%%%%%%%%%%%%
%%%%%%%%%%%%%%%%%%%%%%%%%%%%%%%%%%%%%%%%%%%%% SECTION 3 %%%%%%%%%%%%%%%%%%%%%%%%%%%%%%%%%%%%%%%%%%%%%%%%%%%%%%%%%%%%%%%%%%%%%%%%%%%%
\section{Cosmological Solutions of the Field Equations}
Let us choose such a solution of the field equations $(9)$ \& $(10)$ so that $\tilde{G}\neq0$ and
\begin{equation}\label{13}
\dot{\rho}+3H(\rho+p)=0
\end{equation}
Therefore, the Eq. $(12)$ becomes
\begin{equation}\label{14}
\dot{S}-\frac{(3\tilde{G}+2)\dot{\tilde{G}}}{(1+\tilde{G})\tilde{G}}S+6HS=0
\end{equation}

Here, we have taken $f(Q,T)=\alpha Q+\beta Q^{2}+\gamma T$ which is obtained by replacing $R$ by $Q$ in $f(R,T)$-gravity in which some authors 
have studied the form $f(R,T)=R+\alpha R^{2}+\beta T$ with $R$ as Ricci scalar curvature and $T$ is the trace of energy momentum tensor $T_{ij}$ and here, 
factor $R^{2}$ is added to explain the late time acceleration in expanding universe and this study motivated us to investigate the above form of $f(Q,T)$. 
Now, we obtain $F=f_{Q}=\alpha+2\beta Q$ and $8\pi \tilde{G}=f_{T}=\gamma$ and also, $\dot{\tilde{G}}=0$ and using this in Eq. $(14)$, we get
\begin{equation}\label{15}
\dot{S}+6HS=0
\end{equation}
Integrating Eq. $(15)$, we obtain the Hubble parameter $H$ as

\begin{equation}\label{16}
H=\frac{1}{6}\left[\frac{1}{\beta}\sqrt{\alpha^{2}+\frac{12\beta k}{a^{6}}}-\frac{\alpha}{\beta}\right]^{\frac{1}{2}} 
\end{equation}
where $\alpha$, $\beta$, $\gamma$ are model parameters and $k$ is an integrating constant.\\

The various observational data sets are available in terms of redshift $z$ and it can be easily compared with theoretical studies in terms of redshift. 
Hence, the various parameters need to derive in terms of redshift $z$ and hence, we are used the relationship between the scale-factor and redshift give as
\begin{equation}\label{17}
\frac{a_{0}}{a}=1+z
\end{equation}
where $a_{0}$ is the present value of scale factor and as a standard convention, we will take $a_{0}=1$ throughout the study. Using Eq. (\ref{17}) in (\ref{16}), 
we get 

\begin{equation}\label{18}
H(z)=\frac{1}{6}\left[\frac{1}{\beta}\sqrt{\alpha^{2}+12\beta k (1+z)^{6}}-\frac{\alpha}{\beta}\right]^{\frac{1}{2}} 
\end{equation}
Now, the deceleration parameter $q(t)$ is defined as $q(t)=-1-\frac{\dot{H}}{H^{2}}$ and calculated as

\begin{equation}\label{19}
q(t)=-1+\frac{18\beta k}{a^{6}[\alpha^{2}+\frac{12\beta k}{a^{6}}-\alpha\sqrt{\alpha^{2}+\frac{12\beta k}{a^{6}}}]}
\end{equation}
This in terms of redshift obtained as
	\begin{equation}\label{20}
	q(z)=-1+\frac{18\beta k(1+z)^{6}}{[\alpha^{2}+12\beta k(1+z)^{6}-\alpha\sqrt{\alpha^{2}+12\beta k(1+z)^{6}}]}
	\end{equation}
Now, from Eqs. $(9)$ \& $(10)$ we have calculated the energy density $\rho$ and isotropic pressure $p$ respectively as

\begin{equation}\label{21}
\rho=-\frac{3\alpha}{2(4\pi+\gamma)}H^{2}-\frac{27\beta}{(4\pi+\gamma)}H^{4}-\frac{\gamma k}{4(4\pi+\gamma)(8\pi+\gamma)a^{6}}
\end{equation}

\begin{equation}\label{22}
p=\frac{\alpha}{2(4\pi+\gamma)}H^{2}+\frac{27\beta}{4\pi+\gamma}H^{4}-\frac{k(16\pi+\gamma)}{4(4\pi+\gamma)(8\pi+\gamma)a^{6}}  
\end{equation}
From Eq. $(16)$ one can see that for $\alpha=0$, we can find $H\propto\frac{1}{a^{3/2}}$, $H^{2}\propto\frac{1}{a^{3}}$ and $H^{4}\propto\frac{1}{a^{6}}$ and 
hence, $\rho\propto\frac{1}{a^{6}}$ which shows the existence of stiff matter in the universe. And in late time universe it converted into ordinary matter 
and dark energy.
%%%%%%%%%%%%%%%%%%%%%%%%%%%%%%%%%%%%%%%%%%%%%%%%%%%%%%%%%%%%%%%%%%%%%%%%%%%%%%%%
%%%%%%%%%%%%%%%%%%%%%%%%%%%%%%%%%%%%%%%%%%%%%%%%%%%%%%%%%%%%%SECTION 4%%%%%%%%%%%%%%%%%%%%%%%%%%%%%%%%%%%%%%%%%%%%%%%%%%%%%%%%%%%%%%%%%%%%%%%%%%%%%%%%%%%%
\section{Result Analysis and Discussions}
In the above section, we have obtained the various parameters viz. Hubble function $H(z)$, deceleration parameter $q(z)$, energy density $\rho$ and 
isotropic pressure $p$ in terms of $\alpha$, $\beta$, $\gamma$, $k$, $z$ and $H$ where $\alpha$, $\beta$, $\gamma$, $k$, are called as model parameters. 
To explain the various observable properties of universe, we are to find the best fit values of the model parameters $\alpha$, $\beta$, $\gamma$, $k$ with 
various observational data sets. Therefore, we have found the best fit values of these model parameters for the best fit curve of Hubble function as well 
as apparent magnitude $m(z)$ with observational data sets.
%%%%%%%%%%%%%%%%%%%%%%%%%%%%%%%%%%%%%%%%%%%%%%%%%%%%%%%%Subsection 4.1 %%%%%%%%%%%%%%%%%%%%%%%%%%%%%%%%%%%%%%%%%%%%%%%%%%%%%%%
\subsection{Hubble Parameter}
The Hubble parameter shows the rate of geometrical evolution of the universe (expansion rate) and it can be estimated from observational data sets as in Table-$1$ 
(see \cite{ref77}-\cite{ref83}). The expression for the Hubble parameter $H(z)$ is obtained as in Eq. (\ref{18}). To obtain the best fit values of model parameters 
$\alpha$, $\beta$ and $k$, we have found the best fit curve of Hubble function $H(z)$ with the $29$ observed values of Hubble constant as shown in Table-$1$, 
using the $R^{2}$-test formula as given below:
\begin{equation}\nonumber
R^{2}_{SN}=1-\frac{\sum_{i=1}^{29}[(H_{i})_{ob}-(H_{i})_{th}]^{2}}{\sum_{i=1}^{29}[(H_{i})_{ob}-(H_{i})_{mean}]^{2}}
\end{equation}
The case $R^{2}=1$ shows the exact fit the value of model parameters $\alpha$, $\beta$ and $k$ with observational data sets. Hence, we 
obtain the curve-fitting with greatest value of $R^{2}$. Thus, we have obtained the best fit curve of the Hubble function $(18)$ shown as in figure $1$. 
We have found the best fit values of model parameters $\alpha$, $\beta$ and $k$ as  $\alpha=-3.041$, $\beta=3.686\times10^{-5}$ and $k=1.0\times10^{4}$ 
(see Table-$2$) for maximum $R^{2}=0.9098$ with root mean square error (RMSE) $13.1300$ i.e. $H(z)\pm 13.13$ and their $R^{2}$ values only $9.02\%$ far 
from the best one.\\
%%%%%%%%%%%%%%%%%%%%%%%%%%%%%%%%%%%%%%%%%%%%%%%%%%%%%%%%%%%%%%%%
\begin{table}[H]
	\centering
	{\begin{tabular}{rrrrrrrrrrrrrrrrrrrrrrrrrrrrrrrrrrrrrrrrrrrrrrrrrrrrrrrrr@{}cccccccccccccccccccccccccccccccccccccccccccccccccccccccc@{}}
			\hline\hline 
			$z$  & $H(z)$ & $\sigma_{H}$ & Reference&$z$  & $H(z)$ & $\sigma_{H}$ & Reference\\
			\hline\hline
			$0.070$ \vline & $69$ \vline & $19.6$ \vline &\cite{ref77}\vline& $0.600$ \vline & $87.9$ \vline & $6.1$ \vline &\cite{ref81}\\
			$0.100$ \vline & $69$ \vline & $12$ \vline &\cite{ref78}\vline& $0.680$ \vline & $92$ \vline & $8$ \vline &\cite{ref79}\\
			$0.120$ \vline & $68.6$ \vline & $26.2$ \vline &\cite{ref77}\vline& $0.730$ \vline & $97.3$ \vline & $7$ \vline &\cite{ref81}\\
			$0.170$ \vline & $83$ \vline & $8$ \vline &\cite{ref78}\vline& $0.781$ \vline & $105$ \vline & $12$ \vline &\cite{ref79}\\
			$0.179$ \vline & $75$ \vline & $4$ \vline &\cite{ref79}\vline& $0.875$ \vline & $125$ \vline & $17$ \vline &\cite{ref79}\\
			$0.199$ \vline & $75$ \vline & $5$ \vline &\cite{ref79}\vline& $0.880$ \vline & $90$ \vline & $40$ \vline &\cite{ref82}\\
			$0.200$ \vline & $72.9$ \vline & $29.6$ \vline &\cite{ref77}\vline& $0.900$ \vline & $117$ \vline & $23$ \vline &\cite{ref78}\\
			$0.270$ \vline & $77$ \vline & $14$ \vline &\cite{ref78}\vline& $1.037$ \vline & $154$ \vline & $20$ \vline &\cite{ref79}\\
			$0.280$ \vline & $88.8$ \vline & $36.6$ \vline &\cite{ref77}\vline& $1.300$ \vline & $168$ \vline & $17$ \vline &\cite{ref78}\\
			$0.350$ \vline & $76.3$ \vline & $5.6$ \vline &\cite{ref80}\vline& $1.363$ \vline & $160$ \vline & $33.6$ \vline &\cite{ref78}\\
			$0.352$ \vline & $83$ \vline & $14$ \vline &\cite{ref79}\vline& $1.430$ \vline & $177$ \vline & $18$ \vline &\cite{ref78}\\
			$0.400$ \vline & $95$ \vline & $17$ \vline &\cite{ref78}\vline& $1.530$ \vline & $140$ \vline & $14$ \vline &\cite{ref78}\\
			$0.440$ \vline & $82.6$ \vline & $7.8$ \vline &\cite{ref81}\vline& $1.750$ \vline & $202$ \vline & $40$ \vline &\cite{ref78}\\
			$0.480$ \vline & $97$ \vline & $62$ \vline &\cite{ref82}\vline& $2.300$ \vline & $224$ \vline & $8$ \vline &\cite{ref83}\\
			$0.593$ \vline & $104$ \vline & $13$ \vline &\cite{ref79}\vline& \vline&\vline&\vline&\\
			\hline
	\end{tabular}}
	\caption{Hubble's constant Table.}
\end{table}
%%%%%%%%%%%%%%%%%%%%%%%%%%%%%%%%%%%%%%%%%%%%%%5%%%%%%%%%%%%%%%%%%%%%%%%%%%%%%%%%%%%%%%%%%%%%%%%%%%%%%%%%%%%%%%%%%%%%%%%%%%%%%%%%%%%%
%%%%%%%%%%%%%%%%%%%%%%%%% %%%%%%%%%%%%%%%%%%%%%%%%%%%%%%%%%%%%%%% Figure 1 %%%%%%%%%%%%%%%%%%%%%%%%%%%%%%%%%%%%%%%%%%%%
\begin{figure}[H]
	\centering
	\includegraphics[width=10cm,height=9cm,angle=0]{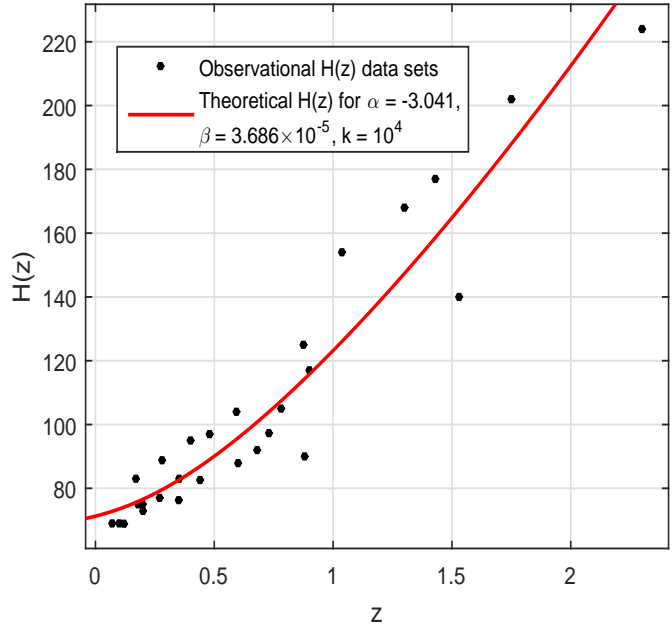}
	\caption{The best fit curve of Hubble function $H(z)$ as in Eq. $(18)$ with observed values of $H(z)$ as shown in Table-$1$. The best fit values of 
	model parameters $\alpha$, $\beta$ and $k$ are obtained with $95\%$ confidence level of bounds.}
\end{figure}
%%%%%%%%%%%%%%%%%%%%%%%%%%%%%%%%%%%%%%%%%%%%%%%%%%%%%%%%%%%%%%%%%%%%%%%%%%%%%%%%%%%%%%%%%%%%%%%%%%%%%%%%%%%%%%%
The Figure $1$ represents the best fit curve of Hubble parameter $H(z)$ for the obtained values of the constants $\alpha$, $\beta$ and $k$. For these values, 
it is observed that the curve is quiet consistent. The present value of Hubble parameter is calculated as $H_{0}=71.26\pm13.13$, which are very close to the 
recent observations \cite{ref86}. \\
%%%%%%%%%%%%%%%%%%%%%%%%%%%%%%%%%%%%%%%%%%%%%%%%%%%%%%%%%%%%%%%%%%%%%%%%%%%%%%%%%%
\begin{table}[H]
	\centering
	{\begin{tabular}{ccccc}
			\hline\hline
			Parameters & $H(z)$ & JLA & SNe Ia \\
			\hline
			$\alpha$ & $-3.041$ & $-2.814$ & $-2.940$\\
			
			$\beta$ & $3.686\times10^{-5}$ & $2.432\times10^{-5}$ & $3.686\times10^{-5}$\\
			
			$k$ & $1.0\times10^{4}$ & $1.0063\times10^{4}$ & $1.0575\times10^{4}$\\
			
			$H_{0}$ & $71.26$ & $54.62$ & $83.53$\\
			
			$R^{2}$ & $0.9098$ & $0.9258$ & $0.9938$\\
			
			RMSE & $13.1300$ & $0.2681$ & $0.2494$\\
			\hline\hline
	\end{tabular}}
	\caption{The best fit values of model parameters $\alpha$, $\beta$ and $k$ for the best fit curve of Hubble function $H(z)$ and apparent magnitude 
	$m(z)$ with different observational data sets ($H(z)$, JLA, SNe Ia) with $95\%$ confidence level of bounds.}
\end{table}
%%%%%%%%%%%%%%%%%%%%%%%%%%%%%%%%%%%%%%%%%%%%%%%%%%%%%%%%%%%%%%%%%%%%%%%%%%%Subsection 4.2 %%%%%%%%%%%%%%%%%%%%%%%%%%%%%%%%%%%%%%%%%%%%%%%%%%%%%%%%%%%%%%%%%%%%%%%%%
\subsection{Luminosity Distance, Apparent Magnitude}
The luminosity distance is measured the total flux of the source of light and is defined as $D_{L}=c(1+z)\int_{0}^{z}\frac{dz}{H(z)}$. Let us define the apparent 
magnitude in terms of $D_{L}$ to obtain the best fit values of model parameters $\alpha$, $\beta$ and $k$ for the best fit curve of $H(z)$ with SNe Ia data sets 
as $m(z)=16.08+5 log_{10}(\frac{H_{0}D_{L}}{0.026 \text{Mpc}})$. To find the best fit curve of apparent magnitude $m(z)$ with SNe Ia data sets we have considered 
$51$ observed data sets of apparent magnitude $m(z)$ from Joint Light curve Analysis (JLA) as in \cite{ref84} and $580$ observed data sets of apparent magnitude 
from union $2.1$ compilation \cite{ref85}. To perform the best curve-fitting of theoretical and observed results we used the following $R^{2}$-test formula:
\begin{equation}\nonumber
R^{2}_{SN}=1-\frac{\sum_{i=1}^{N}[(m_{i})_{ob}-(m_{i})_{th}]^{2}}{\sum_{i=1}^{N}[(m_{i})_{ob}-(m_{i})_{mean}]^{2}}
\end{equation}
where sums are run over $1$ to $N$ and $N$ has values $51$ for JLA data and $580$ for union 2.1 compilation data sets.\\

The ideal case $R^{2}=1$ occurs when the observed data and theoretical function $m(z)$ agree exactly. On the basis of maximum value of $R^{2}$, we get 
the best fit values of $\alpha$, $\beta$ and $k$ for the apparent magnitude $m(z)$ function which is given in Table $2$ for both JLA and SNe Ia data sets. 
The best fit curve of $m(z)$ with observed values of $m(z)$ are shown in figures $2a$ \& $2b$ respectively.\\

From Eq. (\ref{16}) we can find the scale-factor $a(t)$ in terms of Hubble constant as
\begin{equation}\label{23}
a(t)=\left[\frac{12\beta k}{(\alpha+36\beta H^{2})^{2}-\alpha^{2}}\right]^{\frac{1}{6}} 
\end{equation}
One can calculate the value of scale-factor for any value of Hubble constant $H$ using Eq. (\ref{23}) and it can see that as $H\to\infty$ then $a\to0$ which 
shows singularity of the model. Also, from Eq. $(23)$ we can derive the scale-factor in terms of cosmic time as well as in terms of redshift $z$ using 
curve-fitting or cosmographic study. Eq. (\ref{23}) depicts that for $\beta\to0$, $a(t)\to0$ and model shows a point universe and for $\alpha= 0$ the Hubble 
parameter $H\propto a^{-\frac{3}{2}}$ which gives the power-law and exponential cosmology.
%%%%%%%%%%%%%%%%%%%%%%%%%Figure 2 %%%%%%%%%%%%%%%%%%%%%%%%%%%%%%%%%%%%%%%%%%%%
\begin{figure}[H]
	\centering
	a.\includegraphics[width=7cm,height=7cm,angle=0]{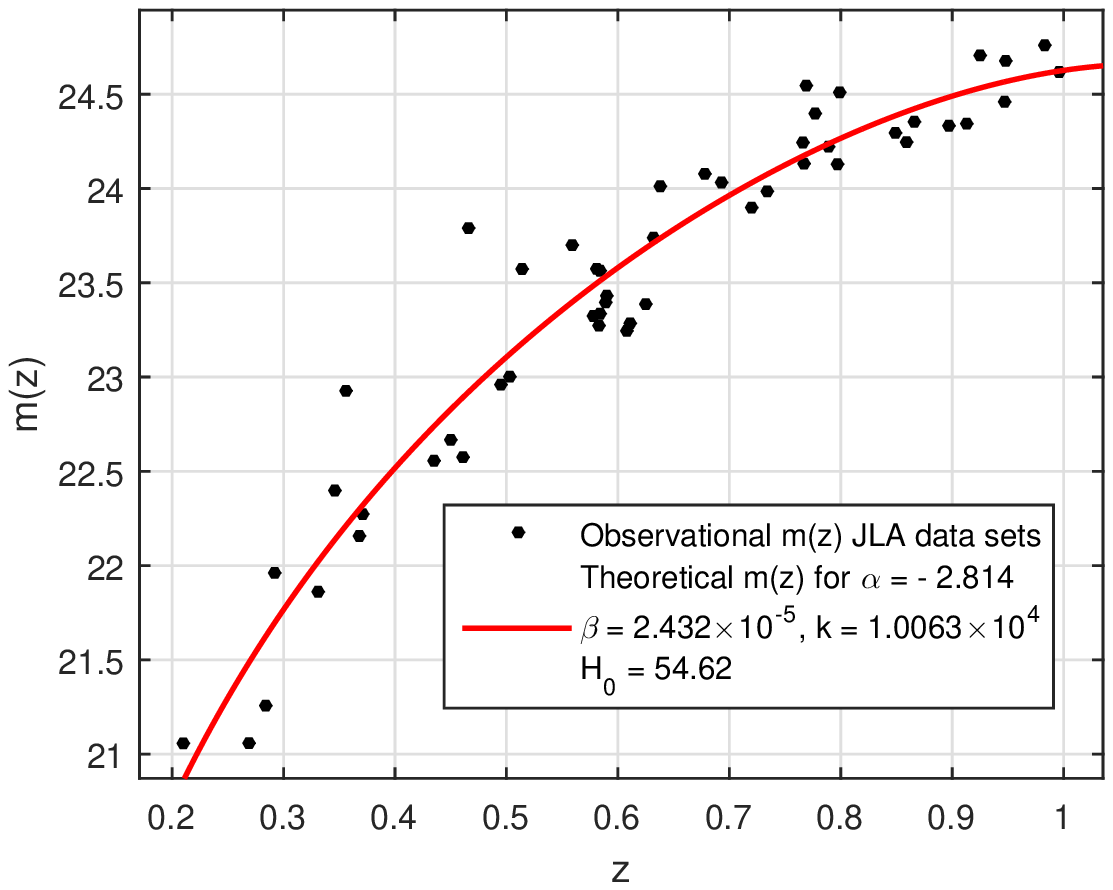}
	b.\includegraphics[width=7cm,height=7cm,angle=0]{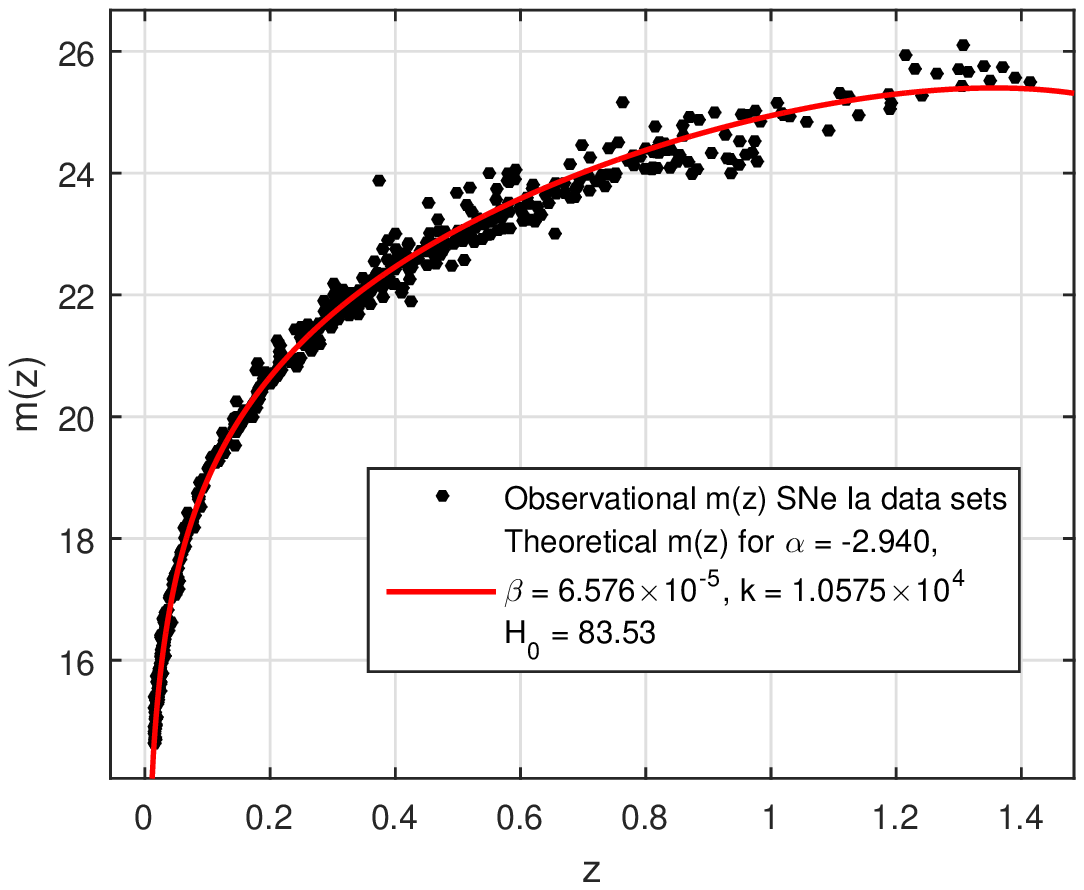}
	\caption{a. The best fit curve of $m(z)$ with observational JLA data sets \cite{ref84} and b. best fit curve of $m(z)$ with union 2.1 compilation 
	of supernovae data sets \cite{ref85}. The best fit values of model parameters $\alpha$, $\beta$ and $k$ are obtained with $95\%$ confidence level of bounds.}
\end{figure}
%%%%%%%%%%%%%%%%%%%%%%%%%%%%%%%%%%%%%%%%%%%%%%%%%%%%%%%%%%%%%%%%%%%%%%%%%Subsection 4.4 %%%%%%%%%%%%%%%%%%%%%%%%%%%%%%%%%%%%%%%%%%%%%%%
\subsection*{Deceleration Parameter}
The deceleration parameter $q(z)$ is the second of the observational parameter that mentions the nature of evolution of geometrical of the universe. 
The expressions for the deceleration parameter (DP) $q(z)$ in terms of scale-factor $a(t)$ and redshift $z$ are shown in Eqs. (\ref{19}) \& (\ref{20}) respectively 
and Figure $3$ shows the variation of DP for the best fit values of model parameters $\alpha$, $\beta$ and $k$ with three data sets. One can see that $q(z)$ is 
an increasing function of redshift $z$ and it shows signature-flipping (transition) point at $z_{t}$ within the range $0.423\leq z_{t}\leq0.668$ 
(as shown in Table-$3$), where $q(z_{t})=0$ and $q(z) < 0$ for $z < z_{t}$ and $q(z) > 0$ for $z > z_{t}$, also, $q(z) \to-1$ as $z\to-1$ and $q(z)$ tends to a 
finite positive value as $z\to\infty$. The present value DP is calculated as $-1\leq q_{0} < 0$ which is mentioned in Table-$3$. The behaviour of DP $q(z)$ over 
redshift $z$ reveals that our universe is undergoing an accelerating phase at present and decelerating in early stages which are very close to the recent 
observations. Thus, our derived model depicts a transit phase model from decelerating to accelerating universe as predicted by the recent observational studies 
\cite {ref1,ref2,ref3,ref4}.\\

From Eq. (\ref{20}) we can find the general expression for the transit value of redshift $z=z_{t}$ by taking $q(z)=0$ and solving for $z$ as:
\begin{equation}\label{24}
z_{t}=\left(\frac{2\alpha^{2}}{3\beta k}\right)^{\frac{1}{6}}-1 
\end{equation}
From Eq. (\ref{24}) one can see that as $\beta\to0$ give $z_{t}\to\infty$ i.e. the model of an ever accelerating universe. Also, we can see that for 
the finite value of model parameters $\alpha$, $\beta$ and $k$ we obtain either an transit phase model or decelerating universe by choosing suitable 
values of the model parameters. From the Table-$3$ we can see that the transition point lies within the range $0.423\leq z_{t}\leq0.668$ 
for the best fit values of model parameters with observational data sets.\\
%%%%%%%%%%%%%%%%%%%%%%%%%%%%%%%%%%%%%%%%%%%%%%%%%%%%%%%%%%%%%%%%%%%%% Figure 3 %%%%%%%%%%%%%%%%%%%%%%%%%%%%%%%%%%%%%%%%%%%%
\begin{figure}[H]
	\centering
	\includegraphics[width=10cm,height=8cm,angle=0]{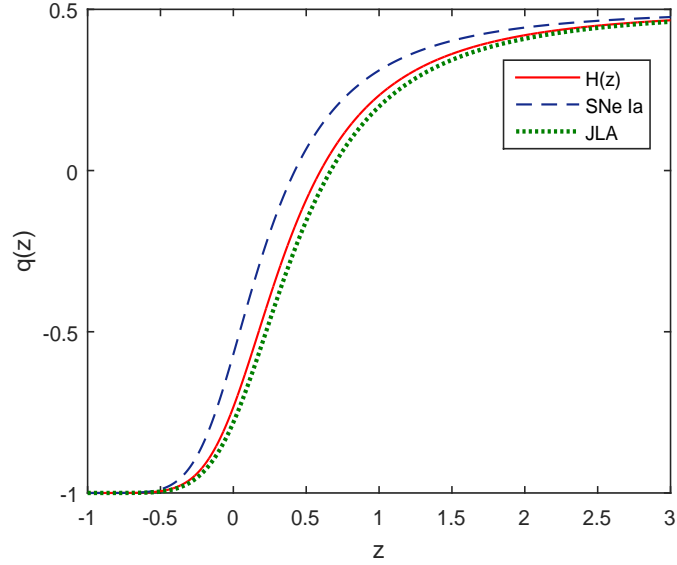}
	\caption{The plot of deceleration parameter $q(z)$ over redshift $z$ for the best fit values $\alpha$, $\beta$ and $k$ which is mentioned in Table-$2$.}
\end{figure}
%%%%%%%%%%%%%%%%%%%%%%%%%%%%%%%%%%%%%%%%%%%%%%%%%%%%%%%%%%%%%%%%%%%%%%%%% Subsection 4.4 %%%%%%%%%%%%%%%%%%%%%%%%%%%%%%%%%%%%%%%%%%%%%%% 
\subsection{Age of the present universe}
The third observational parameter is the age of the present universe $t_{0}$ which is mentioned in Gyrs.
The age of the present Universe is calculated as
\begin{equation}\label{25}
\int_{t_{0}}^{t}dt=\int_{0}^{z}\frac{dt}{dz}dz
\end{equation}
where $t_{0}$ is the present age of the Universe and $t\leq t_{0}$. On integration of left hand side of Eq. $(25)$ and using the relation 
$\frac{dz}{dt}=-(1+z)H(z)$ in right hand side of $(25)$, we obtain
\begin{equation}\label{26}
(t_{0}-t)=\int_{0}^{z}\frac{dz}{(1+z)H(z)}
\end{equation}
where $H(z)$ is the Hubble parameter $H(z)$ (see Eq. $(18)$) with three observational data sets. The plot of cosmic time $(t_{0}-t)H_{0}$ over redshift 
is shown in Figure $4$. The parallel graph of time with $z$-axis shows age of the present universe within the range $11.28\leq t_{0}\leq16.63$ Gyrs which 
is mentioned in Table-$3$. One can conclude that it is very close to age of the present universe calculated in \cite{ref86}.
%%%%%%%%%%%%%%%%%%%%%%%%%%%%%%%%%%%%%%%%%%%%%%%%%%%%%
\begin{table}[H]
	\centering
	{\begin{tabular}{ccccc@rrrrrrrrrrrrrrrrrrrrrrrrrrrrrr}
			\hline\hline
			Parameters & $H_{0}$ & JLA & SNe Ia \\
			\hline
			$q_{0}$ & $-0.7337$ & $-0.7811$ & $-0.5699$\\
			
			$z_{t}$ & $0.599$ & $0.668$ & $0.423$\\
			
			$H_{0}$ & $71.26$ & $54.62$ & $83.53$\\
			
			$t_{0}H_{0}$ & $0.9024$ & $0.9286$ & $0.9636$\\
			
			$t_{0}$ & $12.38$ & $16.63$ & $11.28$\\
			
			\hline\hline
	\end{tabular}}
	\caption{The present values of various parameters obtained for different data sets.}
\end{table}
%%%%%%%%%%%%%%%%%%%%%%%%%Figure 4 %%%%%%%%%%%%%%%%%%%%%%%%%%%%%%%%%%%%%%%%%%%%
\begin{figure}[H]
	\centering
	\includegraphics[width=10cm,height=8cm,angle=0]{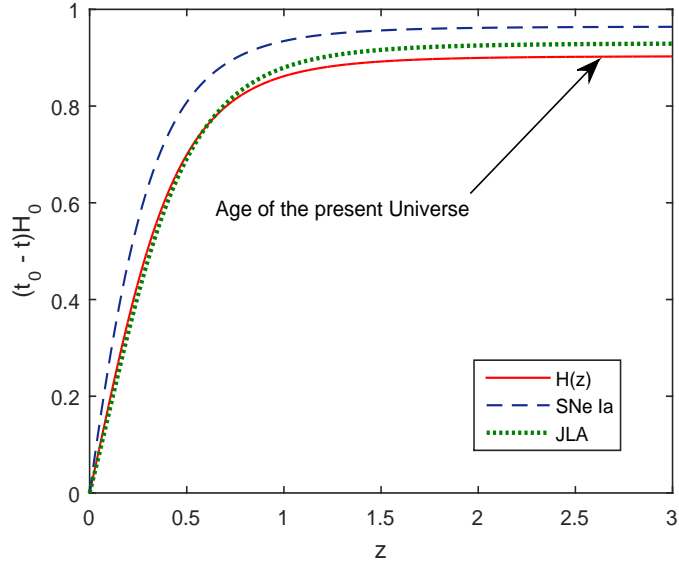}
	\caption{The plot of cosmic time $(t_{0}-t)H_{0}$ over redshift $z$ for the best fit values $\alpha$, $\beta$ and $k$ as given in Table-$2$}
\end{figure}
%%%%%%%%%%%%%%%%%%%%%%%%%%%%%%%%%%%%%%%%%%%%%%%%%%%%%%%%%%%%%%%%%%%%%%%%% 4.5 %%%%%%%%%%%%%%%%%%%%%%%%%%%%%%%%%%%%%%%%%%%%%%%%%%%%%%%%%%%
\subsection*{Energy density $(\rho)$, Isotropic pressure $(p)$ and EoS parameter $(\omega)$}
The expression for energy density $\rho$ is represented by the Eq. $(21)$ and one can see that it has two singularities at 
$\gamma\in\{-4\pi,-8\pi\}$ and $\rho\geq0$ for all values of $\gamma <-8\pi$ and $\rho\leq0$ for $\gamma>-8\pi$. Also, we can find that $\rho$ is 
an increasing function of redshift $z$ with $\gamma < -8\pi$ which shows that as $t\to t_{0}$ ($z\to0$), $\rho$ is decreasing function of redshift 
which is consistent with cosmological studies in observational as well as in theoretical. Eq. $(22)$ represents the expression for isotropic pressure 
$p$ and we can see that the expression of $p$ has two singularities at $\gamma\in\{-4\pi,-8\pi\}$ and $p\leq0$ for all values of $\gamma$ except 
$\gamma=-4\pi, -8\pi$. We can find that the energy density $\rho\geq0$ and the isotropic pressure $p\leq0$ for all values of $\gamma<-8\pi$. Also, one 
can see that the derived model tends to Einstein's GR model  for $k=0$, $\beta=0$ and $\gamma=-8\pi$.\\
%%%%%%%%%%%%%%%%%%%%%%%%%%%%%%%%%%%%%%%%%%%%%%%%%%%%%%%%%%%%% Figure 5 %%%%%%%%%%%%%%%%%%%%%%%%%%%%%%%%%%%%%%%%%%%%
\begin{figure}[H]
	\centering
	a.\includegraphics[width=5cm,height=5cm,angle=0]{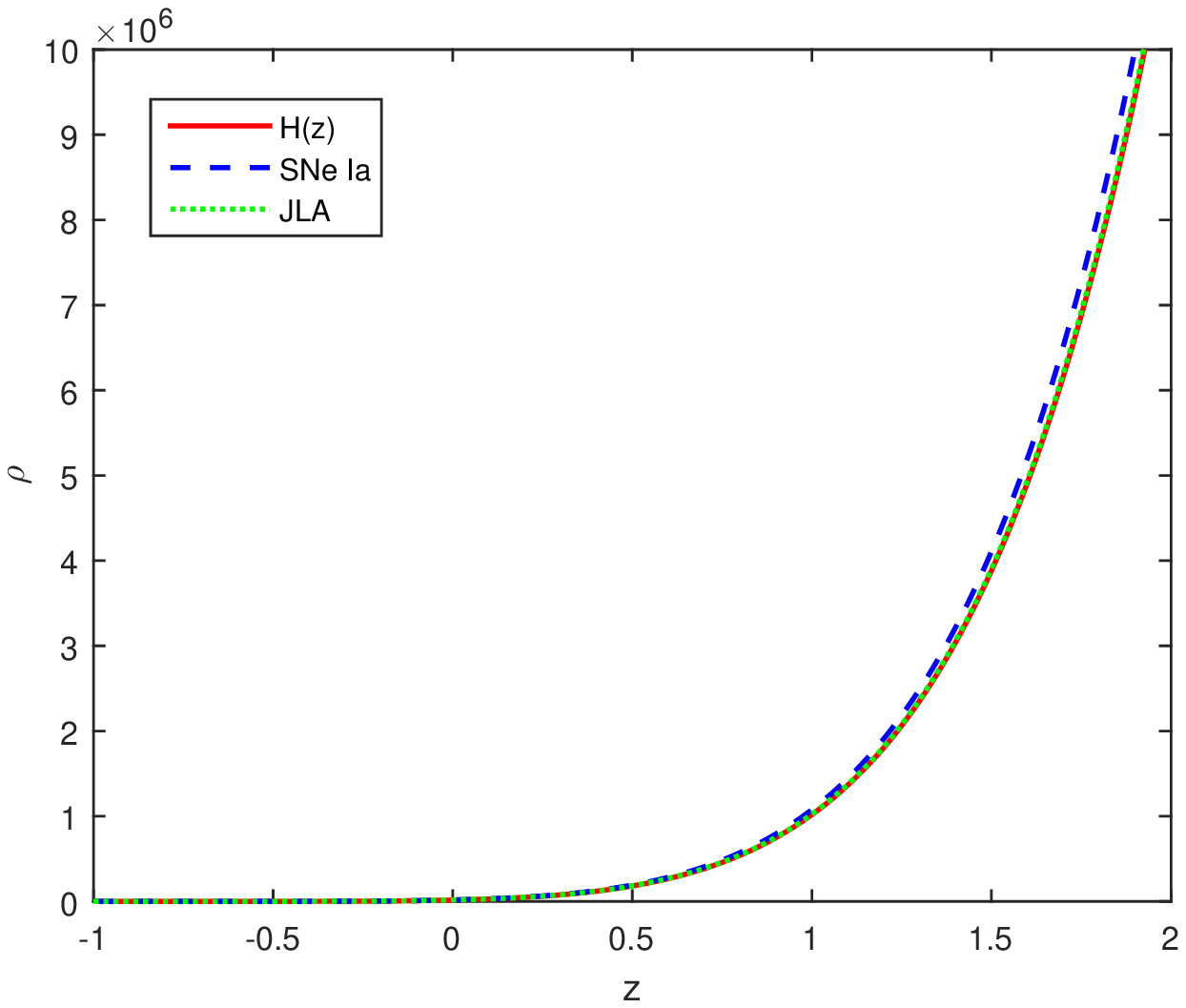}
	b.\includegraphics[width=5cm,height=5cm,angle=0]{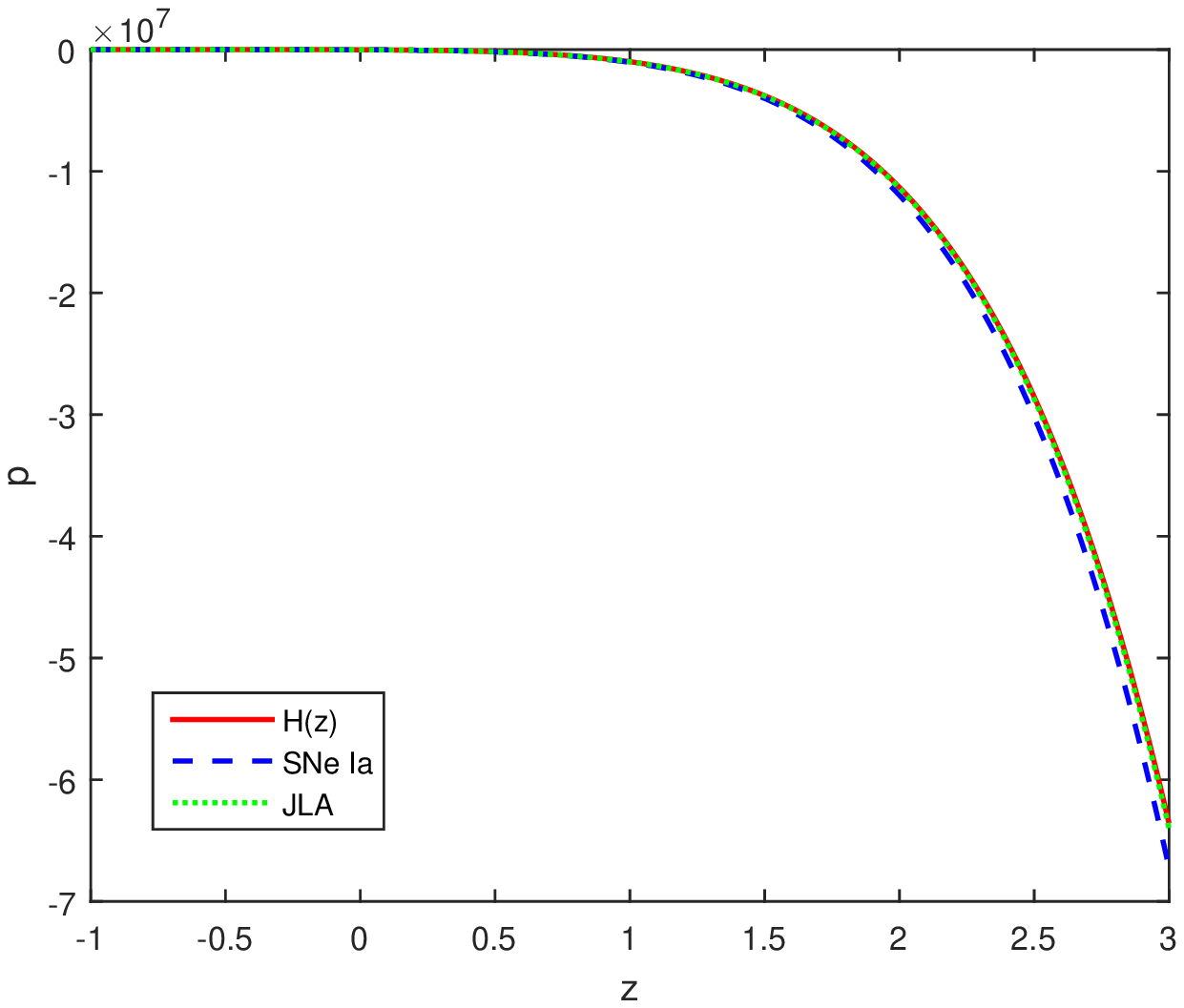}
	c.\includegraphics[width=5cm,height=5cm,angle=0]{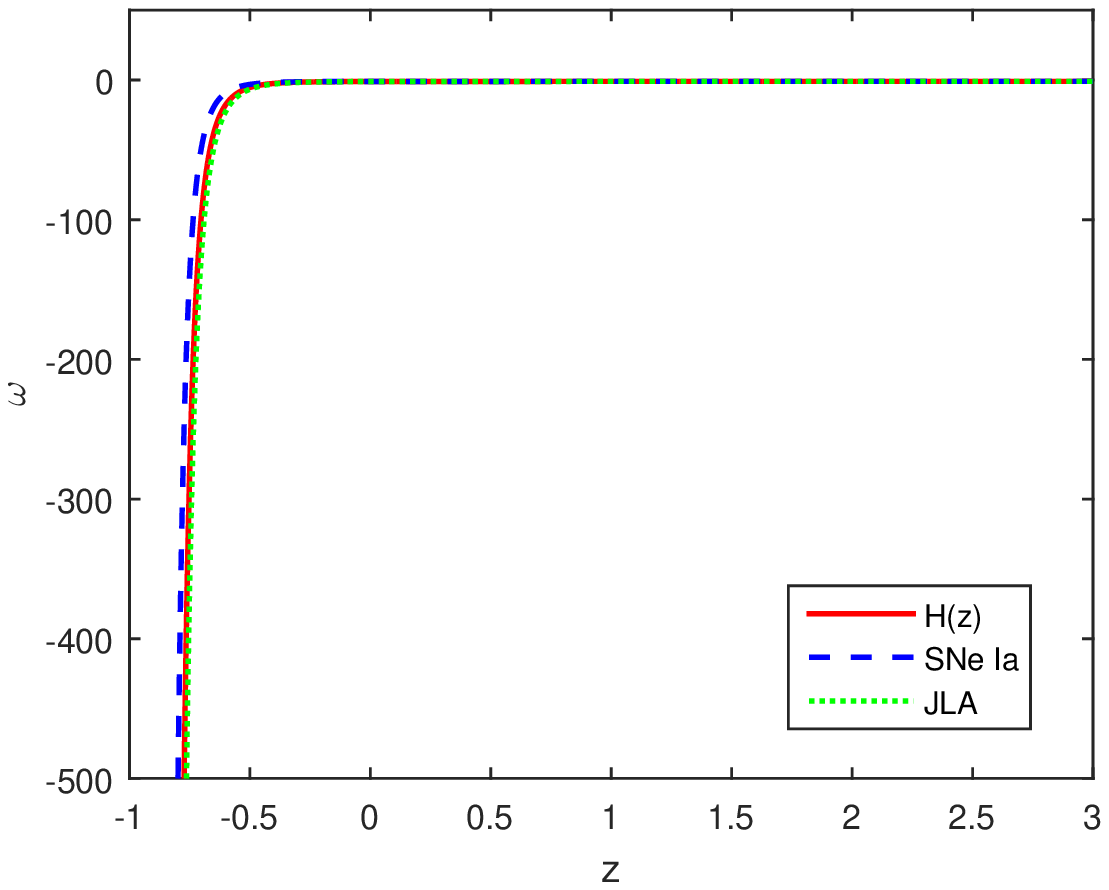}
	\caption{a. The plot of energy density $\rho$ versus $z$, b. the plot of isotropic pressure $p$ versus $z$, and c. the plot of 
	Equation of state parameter (EoS) $\omega$ versus $z$ for the best fit values of $\alpha$, $\beta$ and $k$ mentioned in Table-$2$.}
\end{figure}
%%%%%%%%%%%%%%%%%%%%%%%%%%%%%%%%%%%%%%%%%%%%%%%%%%%%%%%%%%%%%%%%%%%%%%%%%
Figure $5a$ represents the evolution of energy density $\rho$ for $\gamma=-8.1\pi$ and it depicts that for $z\to\infty$, $\rho\to\infty$ 
which shows the big-bang singularity and at late time it tends to zero, while figure $5b$ reveals the behaviour of isotropic pressure $p$ over redshift 
$z$ for $\gamma=-8.1\pi$ and we can see that it is a decreasing function of redshift and $p\to-\infty$ as $z\to\infty$. And figure $5c$ depicts the variation 
of $\omega$ for $\gamma=-8.1\pi$ over redshift $z$ and one can see that it shows $\omega$ is an increasing function of redshift $z$ but $\omega\leq0$. From 
figure $5c$ we can obtain the present value of $\omega$ lies within the range $-1.071\leq\omega\leq-0.96$ for $0\leq z\leq3$ which is agreed with observational 
results in \cite{ref87,ref88,ref89,ref90}. Our model evolves from a quintessence universe to phantom dominated universe and the range of $\omega$ is supported 
to an accelerated universe.\\
%%%%%%%%%%%%%%%%%%%%%%%%%%%%%%%%%%%%%%%%%%%%%%%%%%%%%%%%%%%%%%%%%%%%Subsection 4.6 %%%%%%%%%%%%%%%%%%%%%%%%%%%%%%%%%%%%%%%%%%%%%%%%
\subsection{Statefinder Diagnostic}
In cosmology, we are known two geometrical parameters the Hubble parameter $H=\frac{\dot{a}}{a}$ and the deceleration parameter $q=-\frac{a\ddot{a}}{\dot{a}^{2}}$ 
where $a(t)$ is the scale-factor and these parameters are describe the history of universe. There are another geometrical parameters called as statefinder 
diagnostic proposed in \cite{ref91} which represent the geometric evolution of various stages of dark energy models \cite{ref91,ref92,ref93}. The statefinder 
parameters $r$ and $s$ are defined in terms of scale-factor $a(t)$ respectively as 
\begin{equation}\label{27}
r=\frac{\dddot{a}}{aH^{3}} \hspace{2cm} s=\frac{r-1}{3(q-\frac{1}{2})}
\end{equation}
For the model we obtain statefinder parameters as
\begin{equation}\label{28}
r=1+\frac{3k[1-12\alpha^{2}\beta k\frac{1}{a^{6}}-144\beta^{2}k^{2}\frac{1}{a^{12}}]}{2a^{6}H^{2}\sqrt{\alpha^{2}+\frac{12\beta k}{a^{6}}}}
\end{equation}
\begin{equation}\label{29}
s=\frac{36\beta k[1-12\alpha^{2}\beta k\frac{1}{a^{6}}-144\beta^{2}k^{2}\frac{1}{a^{12}}]}{[12\beta k-a^{6}(\alpha^{2}+\frac{12\beta k}{a^{6}}-
\alpha\sqrt{\alpha^{2}+\frac{12\beta k}{a^{6}}})]}
\end{equation}
%%%%%%%%%%%%%%%%%%%%%%%%%%%%%%%%%%%%%%%%%%%%%%%%%%%%%%%%%%%%%%%%%%%%%% Figure 6 %%%%%%%%%%%%%%%%%%%%%%%%%%%%%%%%%%%%%%%%%%%%
\begin{figure}[H]
	\centering
	  a.\includegraphics[width=7cm,height=7cm,angle=0]{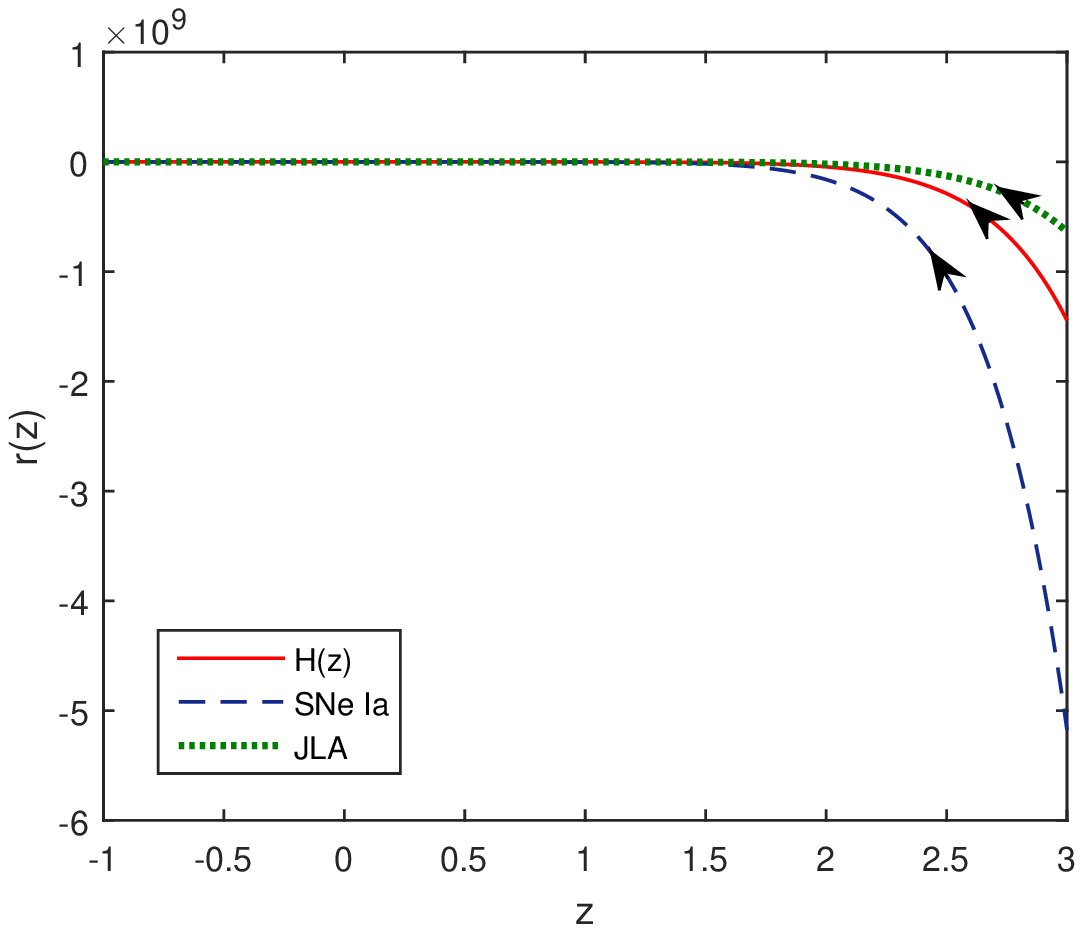}
	  b.\includegraphics[width=7cm,height=7cm,angle=0]{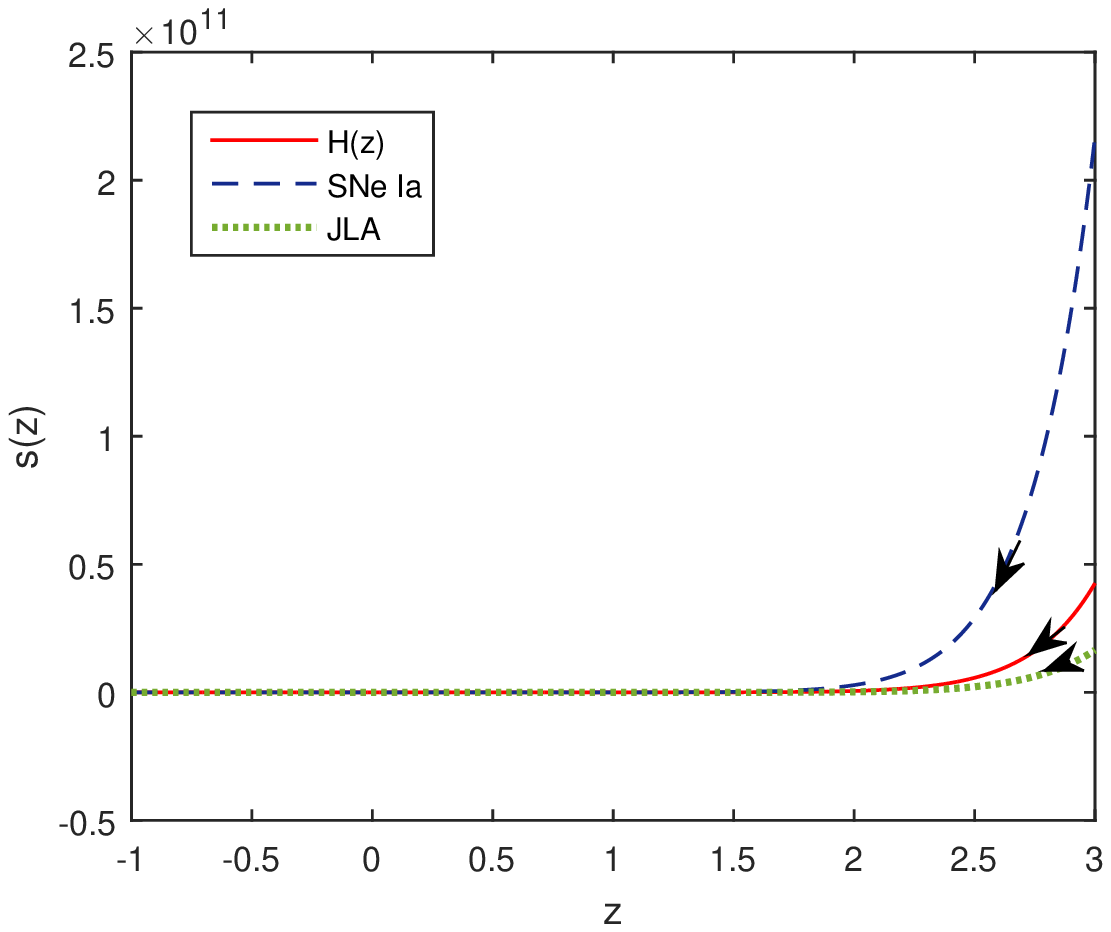}
     \caption{The plot of statefinder parameters $r(z)$, $s(z)$ for the best fit values of $\alpha$, $\beta$ and $k$ mentioned in Table-$2$.}
\end{figure}
%%%%%%%%%%%%%%%%%%%%%%%%%%%%%%%%%%%%%%%%%%%%%%%%%%%%%%%%%%%%%%%%%%%%%%%%%%%%%%%%%%%%%%%%%%%%%%%%%%%%%%%%%%%%%%%%%%%%%%%%%%%%%
%%%%%%%%%%%%%%%%%%%%%%%%%%%%%%%%%%%%%%%%%%%%%%%%%%%%%
\begin{table}[H]
	\centering
	{\begin{tabular}{ccccc@rrrrrrrr}
			\hline\hline
			Parameters & $H(z)$ & JLA & SNe Ia \\
			\hline
			$r_{0}$ & $-46.5099$ & $-19.2762$ & $-180.6409$\\
			
			$s_{0}$ & $38.5101$ & $15.8268$ & $169.7697$\\
			
			$(s,r)_{z\to-1}$ & $(0,1)$ & $(0,1)$ & $(0,1)$\\
			
			\hline\hline
			
	\end{tabular}}
	\caption{The present values of various parameters obtained for different data sets.}
\end{table}
%%%%%%%%%%%%%%%%%%%%%%%%%%%%%%%%%%%%%%%%%%%%%%%%%%%%%%%%%%%%%%%%%%%%%%%%% Figure 7 %%%%%%%%%%%%%%%%%%%%%%%%%%%%%%%%%%%%%%%%%%%%
\begin{figure}[H]
	\centering
	a.\includegraphics[width=7cm,height=7cm,angle=0]{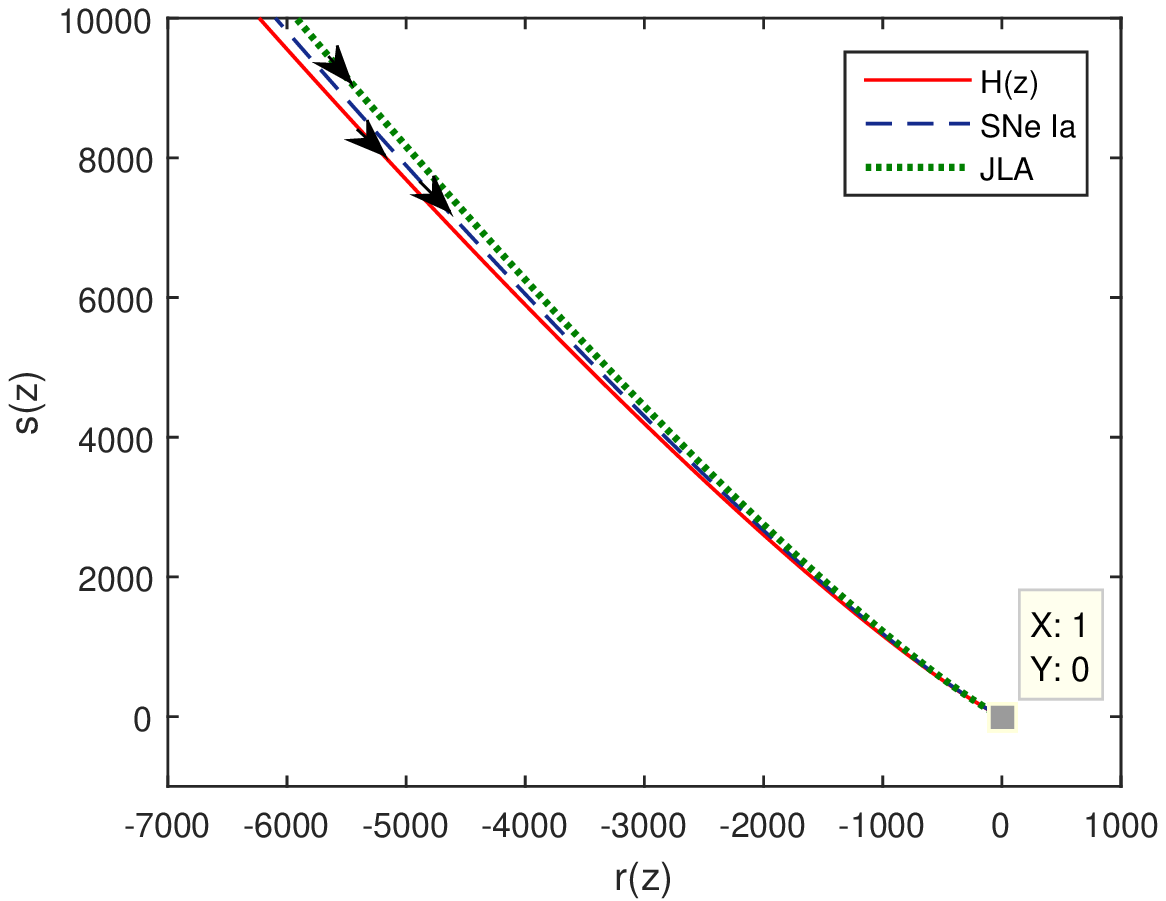}
	b.\includegraphics[width=7cm,height=7cm,angle=0]{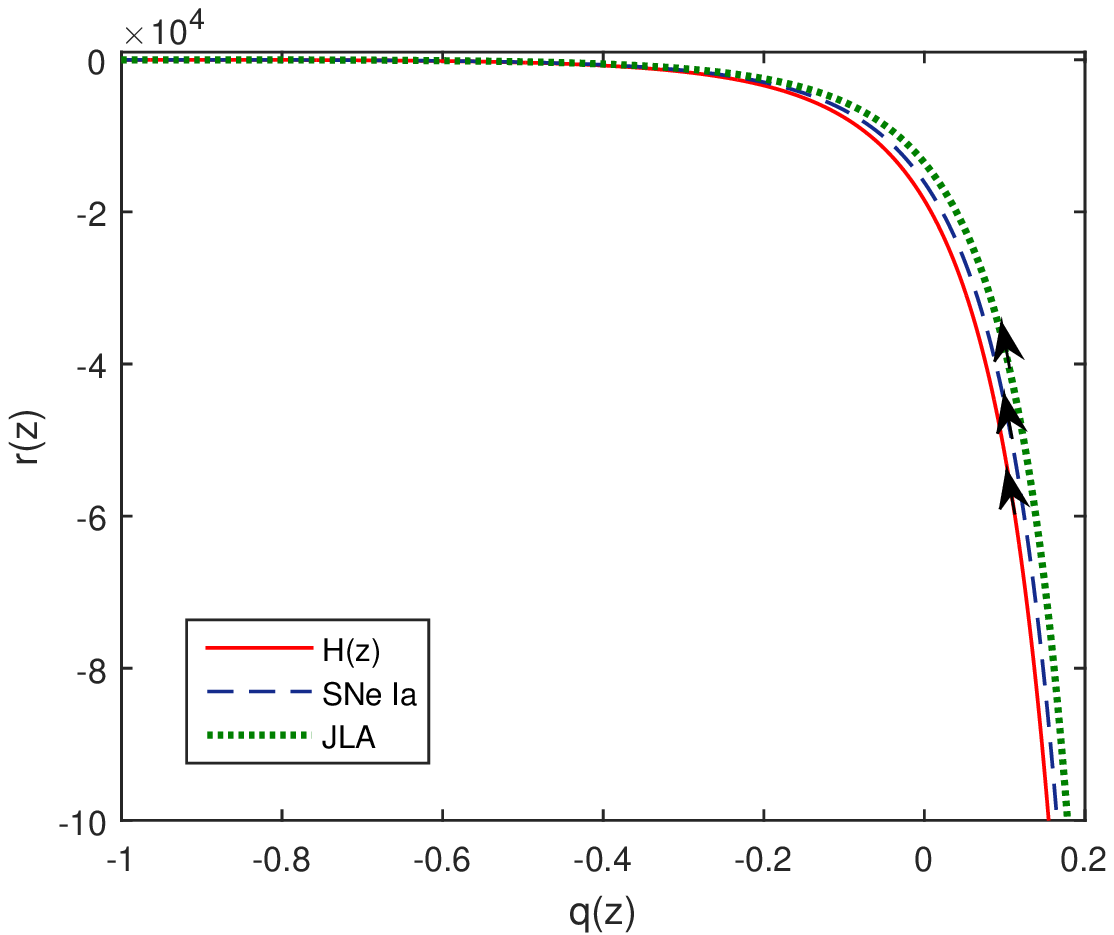}
	\caption{The plot of statefinder parameters $r(z)$, $s(z)$ and $q(z)$, $r(z)$ for the best fit values of $\alpha$, $\beta$ and $k$ mentioned in Table-$2$.}
\end{figure}
%%%%%%%%%%%%%%%%%%%%%%%%%%%%%%%%%%%%%%%%%%%%%%%%%%%%%%%%%%%%%%%%%%%%%%%%%%%%%%%%%%%%%%%%%%%%%%%%%%%%%%%%%%%%%%%%%%%%%
The plot of $r$, $s$ over $z$ are represented in figure $6a$ \& $6b$ and we can calculate the present value of 
$r$ \& $s$ as $-180.6409\leq r_{0}\leq-19.2762$, $15.8268\leq s_{0}\leq169.7697$. Also, one can see that as $z\to-1$ then $r\to1$ and $s\to0$. Now, 
figure $7a$ shows the plot of $s(z)$ over $r(z)$ and the variation of $(s,r)$ represents the various cosmological models \cite{ref91,ref92,ref93} 
and we can see $(s,r)\to(0,1)$ as $z\to-1$ which corresponds to the $\Lambda$CDM and a flat FLRW universe. Figure $7b$ represents the plot of $r(z)$ 
over $q(z)$ and it shows that as $z\to-1$ then $(q,r)\to(-1,1)$ and it confirms that our present and future universe is in accelerating phase of expansion.
%%%%%%%%%%%%%%%%%%%%%%%%%%%%%%%%%%%%%%%%%%%%%%%%%%%%%%%%%%%%%% SECTION 5 %%%%%%%%%%%%%%%%%%%%%%%%%%%%%%%%%%%%%%%%%%%%%%%%%%%%%%%%%%%%%%%%%%%%%%%%%%%%
\section{Conclusions}
In our research, we have taken the function $f(Q,T)$ quadratic in $Q$ and linear in $T$ as $f(Q,T)=\alpha Q+\beta Q^{2}+\gamma T$ where $\alpha$, $\beta$ 
and $\gamma$ are model parameters. We have obtained the various cosmological parameters in Friedmann-Lemaitre-Robertson-walker (FLRW) Universe viz. 
Hubble parameter $H$, deceleration parameter $q$ etc. in terms of scale-factor as well as in terms of redshift $z$ by constraining on energy conservation law. 
For observational constrains on the model, we have obtained the best fit values of model parameters using the available data sets like Hubble data sets $H(z)$, 
Joint Light Curve Analysis (JLA) data sets and union $2.1$ compilation of SNe Ia data sets applying $R^{2}$-test formula. We have calculated the present 
values of various observational parameters $\{t_{0}, q_{0}, H_{0}\}$. These are very close to the standard cosmological models. The main features of our 
model is as follows:
\begin{itemize}
	\item The derived Hubble function is constrained by observational data sets and the present value of Hubble constant is calculated in the range 
	$54.62\leq H_{0}\leq83.53$ which is compatible with \cite{ref86}.
	\item The deceleration parameter $q$ shows the signature-flipping (transition) point within the range $0.423\leq z_{t}\leq0.668$ 
	that depicts our universe has been undergoing in accelerating phase from $t_{z}$ Gyrs ago where $8.85\leq t_{z}\leq14.16$ which 
	is supported by \cite {ref1,ref2,ref3,ref4}.
	\item Our derived model evolves from quintessence universe $-1<\omega<-0.96$ to phantom dominated universe $-1.071<\omega<-1$ 
	passing through vacuum value $\omega=-1$ for $0\leq z\leq3$ which is agreed with the recent observational value obtained in 
	\cite{ref87,ref88,ref89,ref90} and this is the goodness of the derived model in support of late time accelerating scenario of the universe
	\item The age of the present universe is calculated in the range $11.28\leq t_{0}\leq16.63$ compatible with \cite{ref86}.
	\item The statefinder diagnostic is also, analyzed and we obtained $(s,r)\to(0,1)$ as $z\to-1$ (future universe). It shows that our universe 
	model tends to $\Lambda$CDM FLRW a flat universe model in future.
\end{itemize}
Thus, our derived model is much interesting to researchers in this field for further investigations and the compatibility of obtained results with standard 
cosmological models as well as observational data sets shows the viability of our model.
%%%%%%%%%%%%%%%%%%%%%%%%%%%%%%%%%%%%%%%%%%%%%%%%%%%%%%%%%%%%%%%%%%%%%%%%%%%%%%%%%%%%%%%%%%%%%%%%%%%%%%%%%%%%%%%%%%%%%%
% \section*{Acknowledgment}
% The authors are thankful to GLA University Mathura Uttar Pradesh India for providing facilities and support where part of this work is carried out.
%%%%%%%%%%%%%%%%%%%%%%%%%%%%%%%%%%%%%%%%%%%%%%%%%%%%%%%%%%%%%%%%%%%%%%%%%%%%%%%%%%%%%%%%%%%%%
%\newline
%\nonumsection{References}

\end{document}